\documentclass[10pt]{ijnamb}

\def\currentvolume{1}
\def\currentissue{1}
\def\currentyear{2011}
\def\month{January}
\pagespan{1}{18}
\copyrightinfo{2011}{} 

\input{mydef.sty}

\begin{document}

\title[Sensitivity analysis of the early exercise boundary]
{Sensitivity analysis of the early exercise 
\\
boundary for American style of Asian options}

\author[D. \v{S}ev\v{c}ovi\v{c}]{Daniel \v{S}ev\v{c}ovi\v{c}}
\address{
Dept of Applied Mathematics and Statistics, Faculty of Mathematics, Physics and Informatics, Comenius University, 842 48 Bratislava, Slovakia
}
\email{sevcovic@fmph.uniba.sk, martin.taki@gmail.com}

\author[M. Tak\'a\v{c}]{Martin Tak\'a\v{c}}


\commby{Lubin G. Vulkov}

\date{January 15, 2011
}

\thanks{
}

\subjclass[2000]{35K15, 35K55, 90A09, 91B28}

\abstract{
In this paper we analyze American style of floating strike Asian call options belonging to the class of financial derivatives whose payoff diagram depends not only on the underlying asset price but also on the path average of underlying asset prices over some predetermined time interval.
The mathematical model for the option price leads to a  free boundary problem for a parabolic partial differential equation. Applying fixed domain transformation and transformation of variables we develop an efficient numerical algorithm based on a solution to a non-local parabolic partial differential equation for the transformed variable representing the synthesized portfolio. For various types of averaging methods we investigate the dependence of the early exercise boundary on model parameters. 
}

\keywords{Option pricing, American-style Asian options, early exercise boundary, fixed domain transformation}

\maketitle

\section{Introduction}
Asian path dependent options belong to the class of financial derivatives whose payoff diagram depends not only on the underlying asset price but also on the path average of underlying asset prices over some predetermined time interval. Such path dependent options can be often found at commodities markets such as oil, grain trade, etc. At expiration, the payoff diagram of such options is less sensitive with respect to sudden changes of the underlying asset value. Therefore a holder of an Asian option can effectively hedge the risk arising from a sudden price jump close to 
expiry. Typically, the payoff diagram of an Asian path dependent option depends on either arithmetic or geometric average of the spot price of the underlying asset. Such contingent claims can be  used as a financial instrument for hedging highly volatile assets or goods. We refer the reader to references \cite{PW-SH-JD-1995,JCH-1997,LW-YKK-HY-1999,ATH-PLJ-2000,JD-2006,MD-YKK-2006,UW-2006,YKK-2008,BCK-YSO-2004,RW-MCF-2003,VL-2004} discussing qualitative and quantitative aspects of pricing Asian path dependent options. 

In this paper, we focus on a special subclass of Asian options. Namely, we will investigate the so-called average strike Asian call options. At the time of expiry $t=T$, a holder of such an option contract has the right (but not obligation) to purchase the underlying asset for the strike price given as the path average of underlying asset prices. This means that the terminal payoff diagram for such an option has the form:
$V(S,A,T) = \max\{S-A, 0\}$, where $S=S_T$ is the spot price of the underlying asset, $A=A_T$ is the path average of the asset prices $S_t, t\in[0,T],$ over the time interval $[0,T]$ and $T>0$ is the time of maturity. 

\begin{figure}
\centering
\includegraphics[width=7.5truecm]{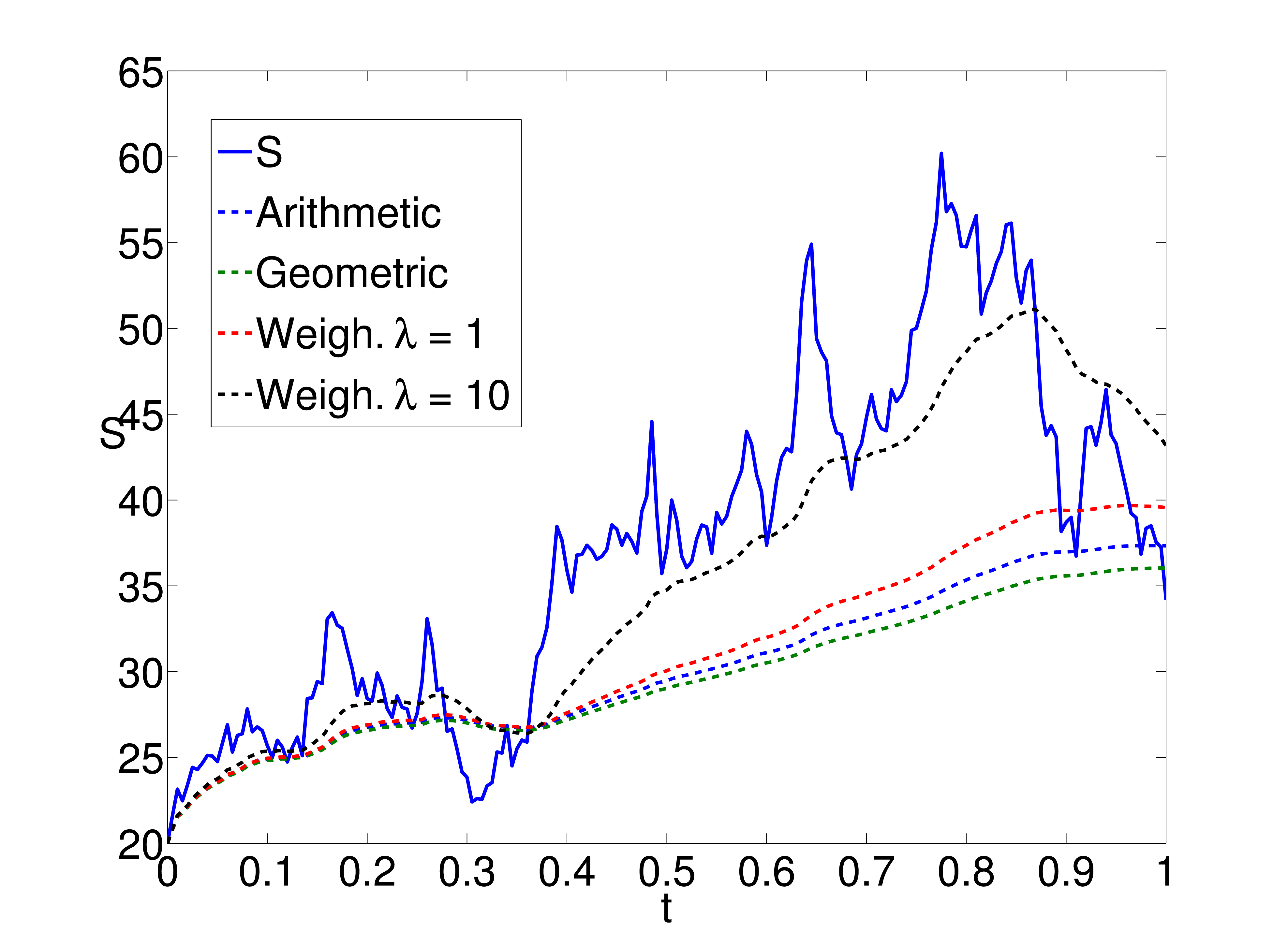}  
\includegraphics[width=7.5truecm]{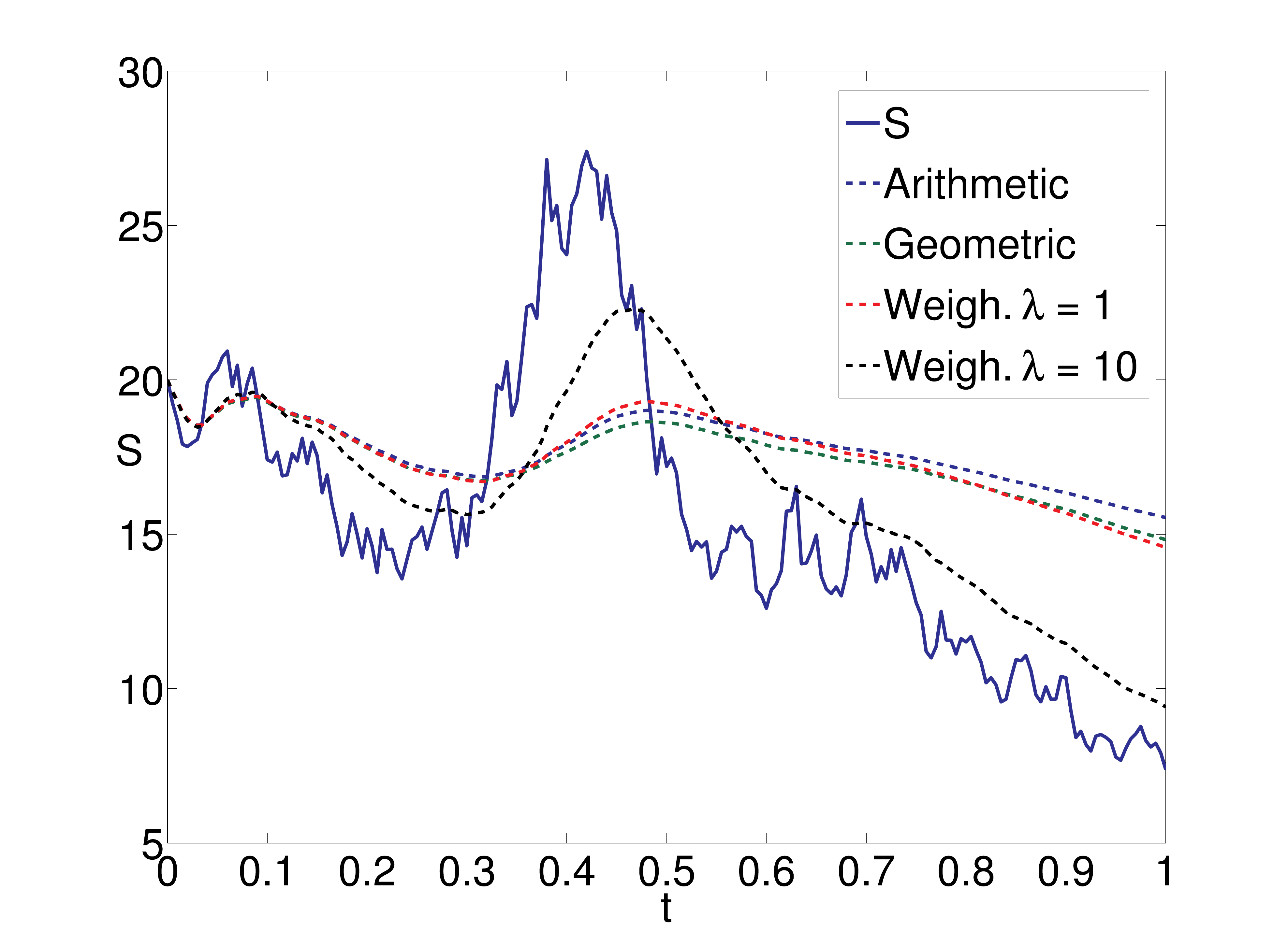} 

\caption{\small Examples of evolution of underlying asset prices (solid lines) and their path averages (dashed lines).}
 \label{fig:asianExample2}
\end{figure}

Concerning the method how the path averaged asset price $A=A_t$ is calculated at a time $t\in[0,T]$,  we can distinguish the following methods of averaging of the path $S_u, u\in[0,t]$:

\begin{itemize}
\item arithmetic averaged options, where the average $A^a_t$ is given by 
\begin{equation}
A^a_t = \frac1t \int_0^t S_\xi \ d\xi,
\label{arit-aver}
\end{equation}

\item  weighted  arithmetic averaged options, with the average $A^{wa}_t$ is given by 
\begin{equation}
A^{wa}_t = \frac{1}{K(t)} \int_0^t a(t-\xi) S_\xi \ d\xi,
\qquad \hbox{where}\ \ K(t)=\int_0^t a(\xi) d\xi,
\label{aritweight-aver}
\end{equation}
and $a$ is an exponential weight function $a(\xi) = \exp(-\lambda\xi)$ with
the averaging factor $\lambda>0$,
\item geometric averaged options, where the average $A^g_t$ is given by 
\begin{equation}
\ln A^g_t = \frac1t \int_0^t \ln S_\xi \ d\xi.
\label{geom-aver}
\end{equation}
\end{itemize}
In Figure \ref{fig:asianExample2} we plot two different sample paths of the underlying asset price (solid lines) and their arithmetic, geometric and weighted arithmetic path averages. In the case of weighted arithmetic averaging with a weight factor $\lambda>0$ we can observe that the average $A^{wa}_t$ approaches the sample path $S_t$ when $\lambda\to +\infty $. On the other hand, the weighted arithmetic average $A^{a}_t$ approaches the arithmetic average when $\lambda\to 0^+ $.

In this paper we are concerned with American style of Asian floating strike options giving its holder the right to exercise it anytime before the obligatory expiration time $t=T$. Our main purpose is to provide a numerical quantitative analysis of the early exercise boundary position for a floating strike Asian call option by means of a solution to the transformed nonlocal parabolic partial differential equation derived by Bokes and \v{S}ev\v{c}ovi\v{c} \cite{sevcovicEarly,sevcovic:nova} for the case of arithmetic averaging. The main goal of this paper is to analyze the dependence and sensitivity of the early exercise boundary with respect to various model parameters. 

The paper is organized as follows. In the next section, we recall the partial differential equation for pricing floating strike Asian options. We also recall the method of dimension reduction of the equation. Next we discuss American style of Asian options and the early exercise boundary. A key tool for derivation of our numerical algorithm is transformation of the reduced equation to a fixed spatial domain. The resulting equation is a nonlocal parabolic PDE with an algebraic constraint between its solution and the free boundary position. In Section 3 we employ the method of finite difference approximation. We discretize our nonlocal equation in space and time. To this end we make use of multiplicative operator splitting method. Results of numerical computations are presented in Section 4. We discuss the impact of the averaging method on the form of the early exercise  boundary. We furthermore analyze the dependence of the early exercise boundary with respect to various model parameters. We furthermore make a comparison of its position for different averaging methods. 

\section{Partial differential equation for pricing average strike Asian options}

In this section we recall the parabolic partial differential equation for pricing Asian options. Derivation of the pricing equation is based on standard assumptions made on stochastic behavior of the underlying asset price $S_t, t\in[0,T]$. Henceforth, we will suppose that $S_t$ follows a geometric Brownian motion, that is,  
\[
dS_t = (r-q) S_t dt + \sigma S_t dB_t,
\]
where $r>0$ is the risk-free interest rate, $q\ge 0$ is the dividend yield on the asset, $B_t, t\in[0,T],$ is the standard Wiener process. Although the aforementioned assumption made on $S_t$ has obvious deficiencies like constancy of the volatility $\sigma$ or normality of the distribution of asset returns, we adopt this assumption throughout the paper. 

Since the payoff diagram $V(S,A,T) = \max\{S-A, 0\}$ depends on both the spot asset price $S$ and the path average $A$ at $t=T$ so does the price $V$ of an Asian option for $0\le t <T$. It means $V$ is a function depending not only on the underlying asset spot price $S$ and time $t$ but also on the average $A$ of the underlying asset price over the interval  $[0,t]$, i.e. $V=V(S,A,t)$. In order to derive the pricing equation for an option price $V$ one has to calculate its differential $dV$ over a time interval with an infinitesimal length $dt$. The price $S=S_t$ as well as its path average $A=A_t$ are stochastic variables and so does the option price  $V_t=V(S_t,A_t,t)$. To calculate the differential $dV$ we have to find a relationship between the differential of the path average $A_t$ and the time $t$. Taking a differential of (\ref{arit-aver}) we obtain the expression: 
\[
\frac{d A^a_t}{dt} = -\frac{1}{t^2}\int_0^t S_\tau d\tau + \frac1t S_t = 
\frac{S_t-A^a_t}{t}
\]
for the case of arithmetic average $A^a_t$. On the other hand, for the case of geometric averaging we obtain from (\ref{geom-aver}):
\[
\frac{1}{A^g_t} \frac{d A^g_t}{dt} = -\frac{1}{t^2}\int_0^t \ln S_\tau d\tau + \frac1t \ln S_t = 
\frac{\ln S_t-\ln A^g_t}{t}.
\]
Similarly, for the exponentially weighted arithmetic average $A^{wa}_t$ with the weight parameter $\lambda>0$  we obtain
\[
\frac{d A^{wa}_t}{dt} = 
\frac{\lambda (S_t-A^{wa}_t)}{1-\exp(-\lambda t)}.
\]
In all averaging methods under consideration, we can conclude the following relation between the differential $dA$ as a function of the stochastic variable $S$ and the differential $dt$.

\begin{equation}
dA = A \, f\left(\frac{S}{A}, t\right) dt,
\label{6kap-dif-priemer}
\end{equation}
where the function $f$ is given by 
\begin{equation}
f(x,t) = \left\{
\begin{matrix}
\displaystyle \frac{x-1}{t}, \hfill & \mbox{for arithmetic averaging},  \hfill
\\
\\
\displaystyle \frac{\lambda (x-1)}{1-\exp(-\lambda t)},  \hfill & \mbox{for exponentially weighted arithmetic averaging},
\\
\\
\displaystyle \frac{\ln(x)}{t}, \hfill & \mbox{for geometric averaging}.\hfill
\end{matrix}
\right.
\end{equation}
It means that the differential $dA$ is a stochastic variable with the leading order term of the order $dt$ (see Kwok \cite{YKK-2008}, Dai \cite{MD-YKK-2006}, \v{S}ev\v{c}ovi\v{c}\cite{sevcovic:nova}). Applying It\=o's lemma (cf. Kwok \cite{YKK-2008}) for the function  $V=V(S,A,t)$ and taking into account (\ref{6kap-dif-priemer}) we conclude the stochastic differential equation for the option price $V$ in the form:
\[
d V = 
\frac{\partial V}{\partial S} dS 
+ \left(
\frac{\partial V}{\partial t} 
+ \frac{\sigma^2}{2} S^2 \frac{{\partial}^2 V}{\partial S^2}
+ \frac{\partial V}{\partial A} A f\biggl(\frac{S}{A}, t\biggr)
\right) dt.
\]
Under the assumption of perfect replicability of the market and nonexistence of arbitrage opportunities (cf. Kwok \cite{YKK-2008}), it can be shown that the risk neutral price $V=V(S,A,t)$ of an Asian call option is a solution to the parabolic partial differential equation
\begin{equation}
 \label{PDR:azijske}
 \frac{\partial V}{\partial t}
 +\frac{\sigma^2}2 S^2 \frac{\partial^2V}{\partial S^2}
 +(r-q)S\frac{\partial V}{\partial S}
 +A f\left(\frac SA,t\right) \frac{\partial V}{\partial A} 
 -r V = 0,
\end{equation}
satisfying the payoff diagram corresponding to the averaged strike Asian option, i.e.
\[
V(S,A,T) = \max\{S-A, 0\},
\]
where $S,A >0,\ t \in (0,T)$. 

It is also well known that the PDE for average strike Asian options allows for a dimension reduction by introducing a new state variable $x$ and the function $W$ defined as follows:
\begin{equation}
x = \frac{S}{A}, \qquad  W(x,\tau) = \frac{1}{A} V(S,A,t),
\label{subst}
\end{equation}
where $\tau = T-t$. After straightforward computations we obtain a parabolic PDE for the function $W(x,\tau)$:
\begin{equation}
 \label{PDR:azijskeRegukovana}
 \frac{\partial W}{\partial \tau}
 +\left[f(x,T-\tau)-r+q \right] x \frac{\partial W}{\partial x}
 - \frac{\sigma^2}2 x^2 \frac{\partial^2 W}{\partial x^2}
 + [r-f(x,T-\tau)] W = 0,
\end{equation}
where $\tau \in (0,T), \ x>0$. A solution $W$ satisfies the initial condition: 
\[
W(x,0 ) = \max\{x-1, 0\}.
\]
A solution $W(x,\tau)$ to the aforementioned equation (\ref{PDR:azijskeRegukovana}) is defined on the fixed spatial interval $0<x<\infty$. The function $V(S,A,t)$ given by $V(S,A,t) = A W(S/A,T-t)$ corresponds to the price of an Asian floating strike path dependent option for the so-called European style of contracts for which the option expires exactly at the time $t=T$.

\subsection{American-style of Asian call options}

In this paper we are concerned with American style of Asian options (cf. \cite{ATH-PLJ-2000,MD-YKK-2006,YKK-2008}). In contrast to European style of options, American style options can be exercised at any time until the obligatory maturity time $t=T$. The holder of such an option has the right to exercise it or to keep it depending on the spot price of the underlying $S_t$ at time $t$ and its history $\{S_u, 0\le u\le t\}$ prior the time $t$. The boundary between ``continuation'' and ``stopping'' regions plays an important role in pricing American-style of options. It can be described by the function 
$(A^*_t, t)\mapsto S^*_t=S_f(A_t,t)$, where $S^*_t$ is the so-called early exercise boundary (cf. \cite{JCH-1997,RG-HEJ-1984,RG-RR-1984,IK-1988,JC-2008,YKK-2008,RAK-JBK-1998,RM-2002,AP-2008}). According to Kwok \cite{YKK-2008}, the set 
\[
\mathcal{E} = \{ (S,A,t) \in [ 0,\infty) \times [ 0,\infty) \times [ 0,T), V(S,A,t) = V(S,A,T)  \}
\]
is the exercise region. For the case of a call option, there exists a early exercise boundary function $S_f = S_f(A,t)$ such that \[
\mathcal{E} = \{ (S,A,t) \in [ 0,\infty) \times [ 0,\infty) \times [ 0,T), S \geq S_f(A,t)  \}.
\]
It means that $S^*_t = S_f(A_t,t)$, where $A_t$ is the path average of underlying asset prices $\{S_u, 0\le u\le t\}$. We can search the early exercise boundary function in a separated form 
\[
S_f(A,t) = A x_f(t).
\] 
For more details we refer the reader to Dai and Kwok \cite{MD-YKK-2006} or Bokes and \v{S}ev\v{c}ovi\v{c} \cite{sevcovicEarly,sevcovic:nova}.
The corresponding spatial  domain for the reduced function $W=W(x,\tau)$ satisfying  (\ref{PDR:azijskeRegukovana}) is therefore given by
\[
0<x<\rho(\tau),\ \tau \in (0,T),
\]
where $\rho(\tau) = x_f(T-\tau)$. From the $C^1$ continuity of $V(S,A,t)$ at  $(S_f(A,t),A,t)$ we conclude that 
\begin{equation}\frac{\partial V}{\partial S}(S_f(A,t),A,t) = 1.
\end{equation}
It follows from the payoff diagram that 
\begin{equation}
V(S_f(A,t),A,t) = S_f(A,t)-A,
\end{equation}
for $A>0$ and $t\in(0,T)$. 
In terms of the new state variable $x$, we conclude the following boundary conditions for the function $W(x,\tau)$:
\begin{align}
 W(0,\tau) &= 0, & W(x,\tau) &=x- 1, & \frac{\partial W}{\partial x} (x,\tau) &= 1,\ \mbox{at} \ x = \rho(\tau),
 \label{eq:derivaciaWatboundary}
\end{align}
for $\tau \in (0,T)$. The initial condition for $W(x,\tau)$ is
\begin{equation}
   W(x,0) = \max\{x-1,0\}, \quad \forall x > 0.
\end{equation}
Equation (\ref{PDR:azijskeRegukovana}) and boundary conditions (\ref{eq:derivaciaWatboundary})  represent a free boundary problem, because the spatial  domain $0<x < \rho(\tau)$  depends on the unknown free boundary function $\rho$ which is a part of a solution of the problem.

\subsection{Fixed domain transformation}

Following ideas of transformation methodology developed by \v{S}ev\v{c}ovi\v{c} in \cite{sevcovic:nova} (see also \cite{sevcovicEarly}) we introduce a new state variable $\xi$ and the transformed function $\Pi=\Pi(\xi,\tau)$ defined as:
\begin{equation}
\xi = \ln\left( \frac{\rho(\tau)}{x} \right),\qquad \Pi(\xi,\tau) = W(x,\tau) -x \frac{\partial W}{\partial x}(x,\tau).
\label{transform}
\end{equation}
After straightforward calculations we obtain that $\Pi(\xi,\tau)$ is a solution to the following linear parabolic equation

\begin{equation}
\frac{\partial \Pi}{\partial \tau} + a(\xi,\tau) \frac{\partial \Pi}{\partial \xi} -\frac{\sigma^2}2\frac{\partial^2 \Pi}{\partial \xi^2} 
+ b(\xi,\tau) \Pi =0, 
\label{finalnaPDR_prePi}
\end{equation}
where 
\[
a(\xi,\tau)= \frac{d}{d\tau}  \ln\rho(\tau) - f(\rho(\tau) e^{-\xi},T-\tau)+r-q- \frac{\sigma^2}2,
\]
\[
b(\xi,\tau) = \left. 
r + x\frac{\partial f}{\partial x}  -f(x,T-\tau)
\right|_{x = \rho e^{-\xi} }. 
\]
For details of derivation of (\ref{finalnaPDR_prePi}) we refer the reader to \cite{sevcovicEarly,sevcovic:nova}. The initial condition for the solution $\Pi(\xi,0)$ is:
\begin{equation}
\label{eq:initialCondition}
\Pi(\xi,0) = \left\{   
     \begin{array}{ll}
      -1, \quad &\xi < \ln \rho(0), \\
      0, \quad &\xi > \ln \rho(0).
     \end{array}
             \right.
\end{equation}
The limiting value $\rho(0)$  of the early exercise boundary at expiry $\tau=0$ (i.e. $t=T$) for the continuous arithmetic average type of an Asian option  has been derived by Dai and Kwok \cite{MD-YKK-2006}. For the geometric average it has been discovered by Wu in  \cite{LW-YKK-HY-1999} (see also Detemple \cite[p. 69]{JD-2006}). For arithmetic weighted averaged floating strike call option it has been derived by Bokes and \v{S}ev\v{c}ovi\v{c} in \cite{sevcovicEarly}. The value $\rho(0)$ is given by:

\begin{equation}
\rho(0) = \left\{
\begin{matrix}
\displaystyle \rho^a(0)=\max\left\{ \frac{1+rT}{1+qT}  ,1\right\}, \hfill & \mbox{arithmetic averaging},  \hfill
\\
\\
\displaystyle \rho^{wa}(0)=\max\left\{ \frac{ \lambda + r (1-e^{-\lambda T}) }{ \lambda +q(1-e^{-\lambda T}  )} ,1\right\},  \hfill & \mbox{weighted arith. averaging},
\\
\\
\displaystyle \rho^g(0)=\max\left\{ \tilde x,1\right\}, \hfill & \mbox{geometric averaging}.\hfill
\end{matrix}
\right.
\label{expr-ro}
\end{equation}
In the case of geometric averaging, the auxiliary number $ \tilde x$ entering the expression of $\rho(0)$ is a unique solution to the transcendental equation:
\begin{equation}
\tilde x q T - r T + \ln(\tilde x) = 0. 
\label{geomrovnica}
\end{equation}

Next we recall the boundary conditions for a solution $\Pi$. 
Taking into account equations (\ref{eq:derivaciaWatboundary}) we end up with the Dirichlet boundary conditions:
\begin{align}
  \Pi(0,\tau) &= -1, & \Pi(\infty,\tau) &= 0.
\end{align}
Since $ \frac{\partial \Pi}{\partial \xi} = x^2 \frac{\partial^2 W}{\partial x^2}$ and $\frac{\partial W}{\partial x}(\rho(\tau),\tau) = 1$
we obtain $\frac{\partial W}{\partial \tau}(\rho(\tau),\tau) = 0$  at   $x=\rho(\tau)$.
In the limit $x \to \rho(\tau)$, assuming the $C^2$ continuity of $\Pi(\xi,\tau)$ up to the boundary $\xi = 0$, we obtain
\[
 x^2 \frac{\partial^2 W}{\partial x^2}(x,\tau) 
 \to \frac{\partial \Pi}{\partial \xi}(0,\tau), 
\qquad 
 x \frac{\partial W}{\partial x} \to \rho(\tau).
\]
Passing to the limit $x\to\rho(\tau)$ in equation (\ref{PDR:azijskeRegukovana}) we obtain the following algebraic constraint between $\rho(\tau)$ and the solution  $\Pi(\xi,\tau)$:
\[
-(r-q) \rho(\tau) -\frac{\sigma^2}2 \frac{\partial \Pi}{\partial\xi}(0,\tau) + r (\rho(\tau) - 1) + f(\rho(\tau),T-\tau) 
= 0.
\] 
Therefore we obtain the following algebraic constraint equation between the free boundary position $\rho(\tau)$ and the partial derivative $\partial_\xi \Pi(0,\tau)$:
\begin{equation}
q \rho(\tau)  -r 
   -\frac{\sigma^2}2 \frac{\partial \Pi}{\partial\xi}(0,\tau) 
   + f(\rho(\tau),T-\tau) = 0.
\label{eq:rovnicaPomocnaIII}
\end{equation}
Notice, that this expression contains term $\frac{\partial \Pi}{\partial\xi}(0,\tau)$ and therefore this is not suitable for numerical scheme, because the whole solution is sensitive of this term. Bokes and \v{S}ev\v{c}ovi\v{c}   \cite{sevcovicEarly} suggested an equivalent form of (\ref{eq:rovnicaPomocnaIII}). They integrated equation (\ref{finalnaPDR_prePi}) with respect to $\xi \in (0,\infty)$. Taking into account boundary conditions for $\Pi(\xi,\tau)$ and $\frac{\partial \Pi}{\partial \xi}(\infty, \tau) = 0$ and using equality (\ref{eq:rovnicaPomocnaIII}) they derived an ODE replacing the algebraic constraint between the free boundary position $\varrho(\tau)$ and $\frac{\partial \Pi}{\partial\xi}(0,\tau)$ of the solution $\Pi$. It has the form:
\begin{eqnarray}
\label{eq:integralnyconstrain}
  0&=& \frac{d}{d\tau} \left( \ln \rho(\tau) 
    + \int_0^{\infty} \Pi(\xi,\tau) d\xi\right)
   +q\rho(\tau) -q -\frac{\sigma^2}{2} 
 \\  & &
   +\int_0^\infty (r-f(\rho(\tau) e^{-\xi},T-\tau))      \Pi(\xi,\tau)  d\xi. \nonumber
\end{eqnarray}

\begin{remark}
Denote by $x^a= (1+rT)/(1+qT)$. Then  $x^a qT -rT + \ln(x^a) < x^a qT -rT + x^a -1 = 0$ provided that $r\not= q$. Since the function $x\mapsto x qT -rT + \ln(x)$ is increasing we have $x^a<\tilde x$ and consequently $\rho^a(0)< \rho^g(0)$. Similarly, for any $\lambda >0$ we have $\rho^{wa}(0) < \rho^a(0)$. In summary, we derived the following inequalities for the initial values of the early exercise boundary
\begin{equation}
\rho^{wa}(0) < \rho^a(0) < \rho^g(0), \quad \hbox{if} \ \ r\not= q.
\label{nerovnosti}
\end{equation}
Clearly, $\rho^{wa}(0) = \rho^a(0) = \rho^g(0)=1$ in the case $r=q$. 

\end{remark}

\begin{remark}
The solution $W=W(x,\tau)$ can be easily calculated from the solution $\Pi$ and the free boundary position $\rho$. With regard to (\ref{transform}) we obtain:
\[
\frac{\partial}{\partial x} \left( x^{-1} W(x,\tau) \right) =
- x^{-2} \Pi\left( \ln\left(\rho(\tau)/x\right), \, \tau \right).
\]
Taking into account the boundary condition $W(x,\tau) = x-1$ at $x=\rho(\tau)$ and integrating the above equation from $x$ to $\rho(\tau)$, we obtain
\[
W(x,\tau)= \frac{x}{\rho(\tau)} 
\left( \rho(\tau) -1   + \int_0^{\ln{\frac{\rho(\tau)}{x}}} {\rm e}^\xi\, \Pi(\xi,\tau) \, d\xi
\right).
\]
Then the option price can be calculated from equality $V(S,A,t) = A W(S/A,T-t)$. 
\end{remark}

\section{Numerical algorithm}

In this section, we make use of the numerical algorithm proposed  by Bokes and \v{S}ev\v{c}ovi\v{c} in \cite{sevcovicEarly,sevcovic:nova} for calculating of the early exercise boundary position $\rho$. We notice that the original algorithm was derived for arithmetic averaged floating strike Asian call option. In this paper, we generalize it for the case of geometric and exponentially weighted averaged floating strike options.

\begin{figure}[htb]
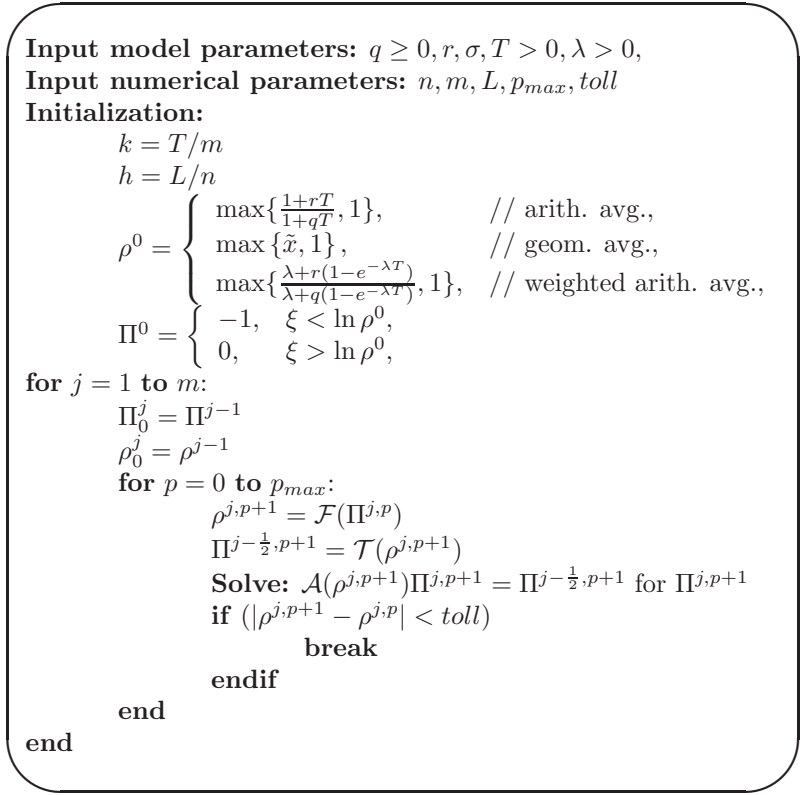


\begin{center}
\Ovalbox{
\vbox{
\vglue3truemm
\begin{tabbing}
\textbf{Input model parameters:} $q\geq 0, r,\sigma,T>0,\lambda>0,$ \\
\textbf{Input numerical parameters:} $n,m, L, p_{max}, toll$ \\
~ \hspace{1cm} \= ~ \hspace{1cm} \=~ \hspace{1cm} \=~ \hspace{1cm} \= \kill 
\textbf{Initialization:} \\
 \> $k=T / m$ \\
 \> $h=L / n$ \\
 \> $\rho^0 = \left\{\begin{array}{ll} 
       \max\{ \frac{1+rT}{1+qT}  ,1\}, & \mbox{// arith.  avg.,  }\\
        \max\left\{ \tilde x,1\right\}
       , & \mbox{// geom. avg.,  }\\
       \max\{ \frac{ \lambda + r (1-e^{-\lambda T}) }{ \lambda +q(1-e^{-\lambda T}  )} ,1\}, & \mbox{// weighted arith. avg.,  }
                     \end{array}   \right.$ \\
 \> ${ \Pi}^0 =  \left\{\begin{array}{ll} 
                           -1, & \xi < \ln \rho^0,\\
			   0, & \xi > \ln \rho^0,
                          \end{array} \right.$ \\
\textbf{for} $j=1$ \textbf{to} $m$: $\quad$   \\
    \> ${ \Pi}^j_0 = { \Pi}^{j-1}$ \\
    \> $\rho^j_0 = \rho^{j-1}$ \\
    \> \textbf{for} $p=0$ \textbf{to} $p_{max}$: $\quad$   \\
    \>    \> $\rho^{j,p+1} = \mathcal{F}({ \Pi}^{j,p})$ \\
    \>    \> ${ \Pi}^{j-\frac12, p+1} = \mathcal{T}( \rho^{j, p+1})$ \\
    \>    \> \textbf{Solve:} $\mathcal{A}( \rho^{j, p+1}){ \Pi}^{j, p+1} = { \Pi}^{j-\frac12, p+1}$ for ${ \Pi}^{j, p+1}$\\
    \>    \> \textbf{if} ($|\rho^{j, p+1} - \rho^{j, p}  | < toll$)    \\ 
    \>    \>     \> \textbf{break} \\
    \>    \> \textbf{endif} \\ 
    \> \textbf{end} \\
\textbf{end}
\end{tabbing}
\vglue3truemm
}}
\end{center}

\caption{A flowchart of the numerical algorithm. Input model parameters: $r$ is the interest rate, $q$ is the dividend yield, $T$ expiration time, $n$ is the number of spatial grid points, $m$ is the time step, $L = 3$, $\lambda$ is the weight parameter.}
\label{flowchart}
\end{figure}

The algorithm is based on a finite difference discretization in space and time variables. We restrict the spatial  domain to a finite interval $\xi \in (0,L)$,  where $L\gg 1$ is sufficiently large. For practical purposes, it is sufficient to take $L\approx 3$. Let $k=\frac Tm>0$ is a time discretization step and $h=\frac Ln > 0$ is a spatial discretization step. We denote by $\Pi^j$  time discretization of $\Pi(\xi,\tau_j)$ and $\rho^j=\rho(\tau_j)$, where $\tau_j = jk$. By $\Pi_i^j$ we denote full space-time approximation of the value $\Pi(\xi_i, \tau_j)$.
Then the Euler backward in time finite difference approximation of (\ref{finalnaPDR_prePi}) reads as follows:
\begin{eqnarray*}
0 &=&
\frac{\Pi^j-\Pi^{j-1}}{k}
+c^j \frac{\partial \Pi^j}{\partial \xi}
-\left(\frac{\sigma^2}2
   + f( \rho^j e^{-\xi},T-\tau)
   \right)\frac{\partial \Pi^j}{\partial \xi}
-\frac{\sigma^2}2\frac{\partial^2 \Pi}{\partial \xi^2} \\
 & &
 +\left[ 
      \left. r + x\frac{\partial f}{\partial x} 
 -f(x,T-\tau)
      \right|_{x = \rho^j e^{-\xi}}   \right]\Pi^j,
\end{eqnarray*}
where $c^j=c(\tau_j)$ and  $c(\tau) = \frac{d}{d\tau} \ln\rho(\tau) + r - q$. We prescribe Dirichlet boundary conditions at $\xi=0$ and $\xi = L$ for the  function $\Pi^j$. As for the initial condition, we use the vector $\Pi^0$ where $ \Pi^0_i = \Pi(\xi_i,0)$. Following \cite{sevcovicEarly}, we make use of the operator splitting method to the above problem  by 
introducing an auxiliary intermediate step $\Pi^{j-\frac12}$ that splits the problem into two parts:
\begin{itemize}
 \item Convective  part 
 \begin{equation}
 \frac{\Pi^{j-\frac12}-\Pi^{j-1}}{k} 
   + c^j  \frac{\partial \Pi^{j-\frac12}}{\partial \xi} = 0, \label{eq:convective}
\end{equation}
 \item Diffusive  part
\begin{eqnarray}
 0 &=&\frac{\Pi^j-\Pi^{j-\frac12}}{k}
-\left(\frac{\sigma^2}2 \label{eq:diffusive}
   + f( \rho^j e^{-\xi},T-\tau)
   \right)\frac{\partial \Pi^j}{\partial \xi}
-\frac{\sigma^2}2\frac{\partial^2 \Pi^j}{\partial \xi^2} \\
 & &
 +\left[  \nonumber
      \left. r + x\frac{\partial f}{\partial x} 
 -f(x,T-\tau)
      \right|_{x = \rho^j e^{-\xi}}   \right]\Pi^j.
\end{eqnarray}
\end{itemize}
A solution $\Pi^{j-\frac12}$ to equation (\ref{eq:convective}) can be approximated by the explicit solution to the transport equation 
\[
\frac{\partial \tilde\Pi}{\partial \tau}
+ c(\tau) 
 \frac{\partial \tilde\Pi}{\partial \xi}
  = 0, 
\]
for $\xi>0$ and $\tau \in (\tau_{j-1},\tau_j ]$ satisfying the initial condition $\tilde \Pi(\xi, \tau_{j-1}) = \Pi^{j-1}(\xi)$ and the boundary condition $\tilde \Pi(0,\tau)=-1$. After some computations (for further details see e.g. Bokes and \v{S}ev\v{c}ovi\v{c} \cite{sevcovicEarly}) we end up with the following solution:
\begin{equation}
  \Pi_i^{j-\frac12} \label{eq:operatorT}
  =
  \left\{
  \begin{array}{ll}
   \Pi^{j-1}(\nu_i), &  \mbox{if}\ 
   \nu_i = \xi_i + \ln \frac{\rho^{j-1}}{\rho^j} - (r-q)k > 0,\\
   -1, & \mbox{otherwise}.
  \end{array}
  \right.
\end{equation}
In order to derive full space--time discretization scheme we make use of the central finite difference approximation of equation (\ref{eq:diffusive}). We obtain 
\begin{eqnarray*}
  0 &=&
  \frac{\Pi_i^j-\Pi_i^{j-\frac12}}{k}
  +\left[  \nonumber
      \left. r + x\frac{\partial f}{\partial x} 
 -f(x,T-\tau)
      \right|_{x = \rho^j e^{-\xi_i}}   \right]\Pi_i^j \\
 & &
-\left(\frac{\sigma^2}2 
   + f( \rho^j e^{-\xi_i},T-\tau)
   \right)\frac{\Pi^j_{i+1}-\Pi^j_{i-1}}{2h}
-\frac{\sigma^2}2
\frac{ \Pi^j_{i+1}-2\Pi^j_{i}+\Pi^j_{i-1} }{h^2}. \\
\end{eqnarray*}
Hence the vector $\Pi^j$ is a solution of a tridiagonal system of linear equations
\begin{equation}
\label{eq:operatorA}
 \alpha_i^j \Pi^j_{i-1}+ \beta_i^j \Pi_i^j
 + \gamma_i^j \Pi^j_{i+1} = \Pi_i^{j-\frac12},
\end{equation}
for $i=1, 2, \dots, n$, where
\begin{eqnarray}
\nonumber
 \alpha^j_i (\rho^j) &=& -\frac{k}{2h^2}\sigma^2 
 + \frac{k}{2h} \left( \frac{\sigma^2}2+ f(\rho^j e^{-\xi_i},T-\tau_j) \right),\\
\nonumber
\beta^j_i (\rho^j) &=& 1+ b(\xi_i,T-\tau_j) k - (\alpha_i^j + \gamma_i^j),  \\
 \gamma^j_i (\rho^j) &=& -\frac{k}{2h^2}\sigma^2 
 \nonumber
 - \frac{k}{2h} \left( \frac{\sigma^2}2+ f(\rho^j e^{-\xi_i},T-\tau_j) \right).
\end{eqnarray}
Boundary conditions for $\Pi^j$ are given by: $\Pi_0^j = -1,\ \Pi_n^j = 0$, for $j=1, 2, \dots, m$.
The initial condition for $\Pi^0$ is given by equations (\ref{eq:initialCondition}) and (\ref{expr-ro}). In order to determine the free boundary position
we take equation (\ref{eq:integralnyconstrain}) into account. Applying the forward finite difference approximation we obtain
\begin{eqnarray}
\ln \rho^j &=& \ln \rho^{j-1}  \label{eq:operatorF}
+ \int_0^\infty \Pi^{j-1}(\xi) d\xi
- \int_0^\infty \Pi^{j}(\xi) d\xi
\\
 & &\nonumber
+ k\left( q+\frac{\sigma^2}{2} -q \rho^{j-1} - \int_0^\infty \left( 
r - f( \rho^{j-1} e^{-\xi} ,T-\tau_j ) \right) \Pi^{j}(\xi) d\xi   \right).
\end{eqnarray}
As for the approximation of the integral $ \int_0^\infty \Pi^{j}(\xi) d\xi$, we use 
the trapezoid quadrature method. If we rewrite equations into the operator form then we obtain the following system of nonlinear algebraic equation for the unknown vector $\Pi^j$ and the free boundary position $\rho^j$ at the time $\tau_j$:
\begin{equation}
\rho^j = \mathcal{F}({\Pi}^j),
\quad
\Pi^{j-\frac12} = \mathcal{T}( \rho^j),
\quad
\mathcal{A}( \rho^j) \Pi^j = \Pi^{j-\frac12},
\label{eq:operators}
\end{equation}
where $\mathcal{T}( \rho^j)$ is a solution of the transport equation given by (\ref{eq:operatorT}),
$\mathcal{A}( \rho^j)$ is a tridiagonal matrix given by (\ref{eq:operatorA}) and
$\ln \mathcal{F}({\Pi}^j)$ is
 right side of equation
 (\ref{eq:operatorF}).
System of equations (\ref{eq:operators}) can be effectively solved by means of successive iterations
procedure. For $j\geq 1$, we set $\Pi^{j,0} = \Pi^{j-1}$ and $\rho^{j,0} = \rho^{j-1}$. Then $(p+1)$-th approximation of $\Pi^j$ and $\rho^j$ is a solution of following system:
\begin{eqnarray}
\rho^{j,p+1} &=&  \mathcal{F}({ \Pi}^{j,p}), \label{eq:1}\\
\Pi^{j-\frac12,p+1} &=&  \mathcal{T}( \rho^{j,p+1}), \label{eq:2}\\
\mathcal{A}( \rho^{j,p+1}) \Pi^{j,p+1} &=& \Pi^{j-\frac12,p+1}. \label{eq:3}
\end{eqnarray}
We repeat the above iteration procedure for $p=1, ..., p_{max}$ until the prescribed tolerance $|\rho^{j,p+1}-\rho^{j,p}| <toll$ is achieved (see Table~\ref{flowchart}).

\section{Numerical results}

The aim of this section is to present various computational examples of calculation of the early exercise boundary position $\rho$ for varying model parameters and averaging methods. In all examples discussed in this section we use the following numerical parameters: $m=10\ 000$, $n=300$, $L=3$, $p_{\max}=500$, $toll=10^{-8}$. In graphical plots we display the position of the early exercise $\rho$ only.

\subsection{Free boundary for floating strike call options} 

\subsubsection{Arithmetic averaged floating strike call options}

\begin{figure}
 \centering
 \includegraphics[width=10truecm]{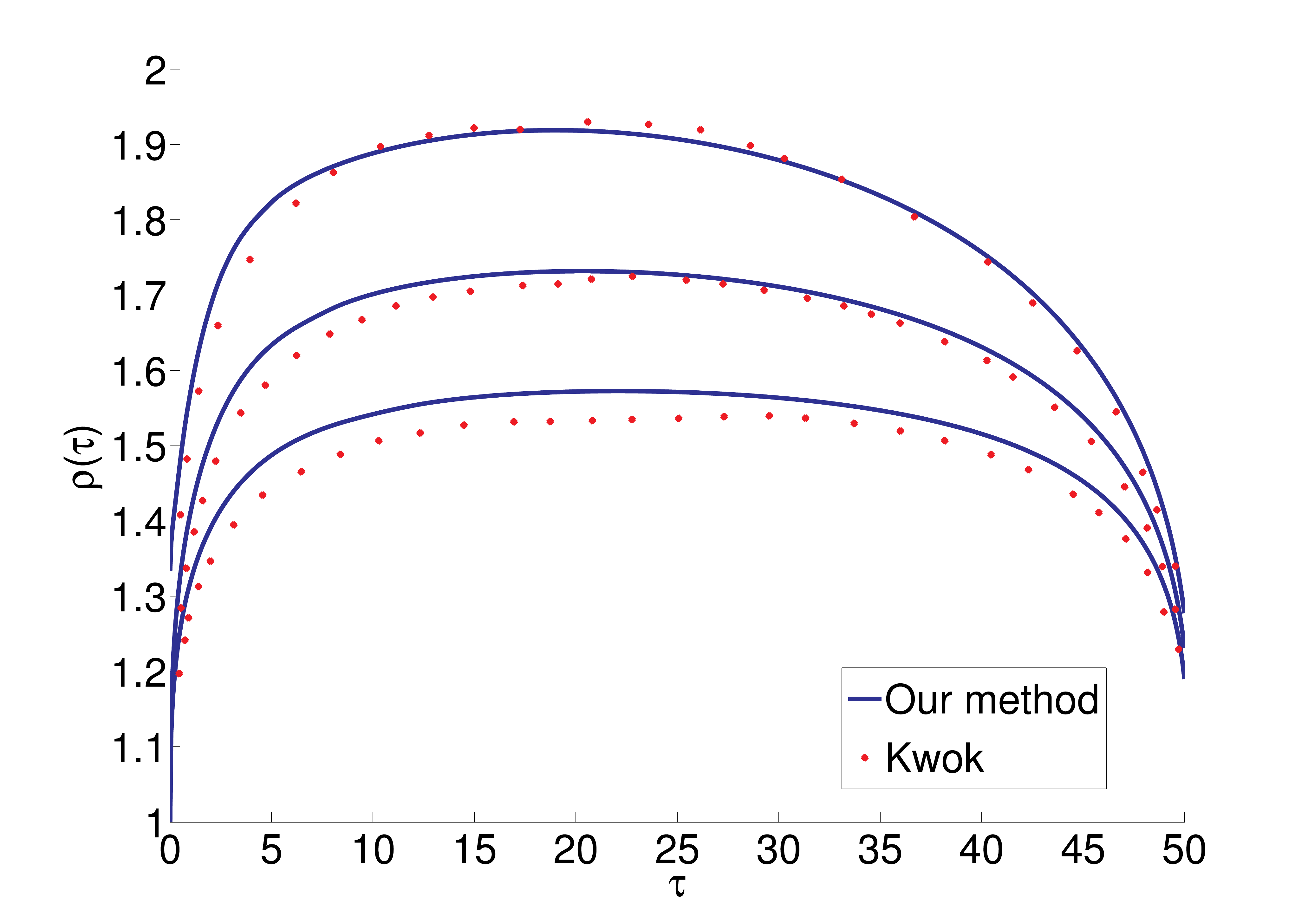}
 \caption{A comparison position of the free boundary position for various $r = 0.02, 0.04, 0.06$. We also compare our results with the method due to Dai and Kwok \cite{MD-YKK-2006}.}
 \label{fig:sevcoviccompare}
\end{figure}

In Figure \ref{fig:sevcoviccompare} we compare the free boundary position $\rho$ for various interest rates $r=0.02, 0.04, 0.06$. Other model parameters are: $T=50, \sigma = 0.2, q=0.04$.
We also compare the free boundary position computed by the algorithm described in Section 3 (blue line)
and computational results obtained by Dai and Kwok in \cite{MD-YKK-2006} (red dots).

\begin{figure}
 \centering
 \includegraphics[width=8truecm]{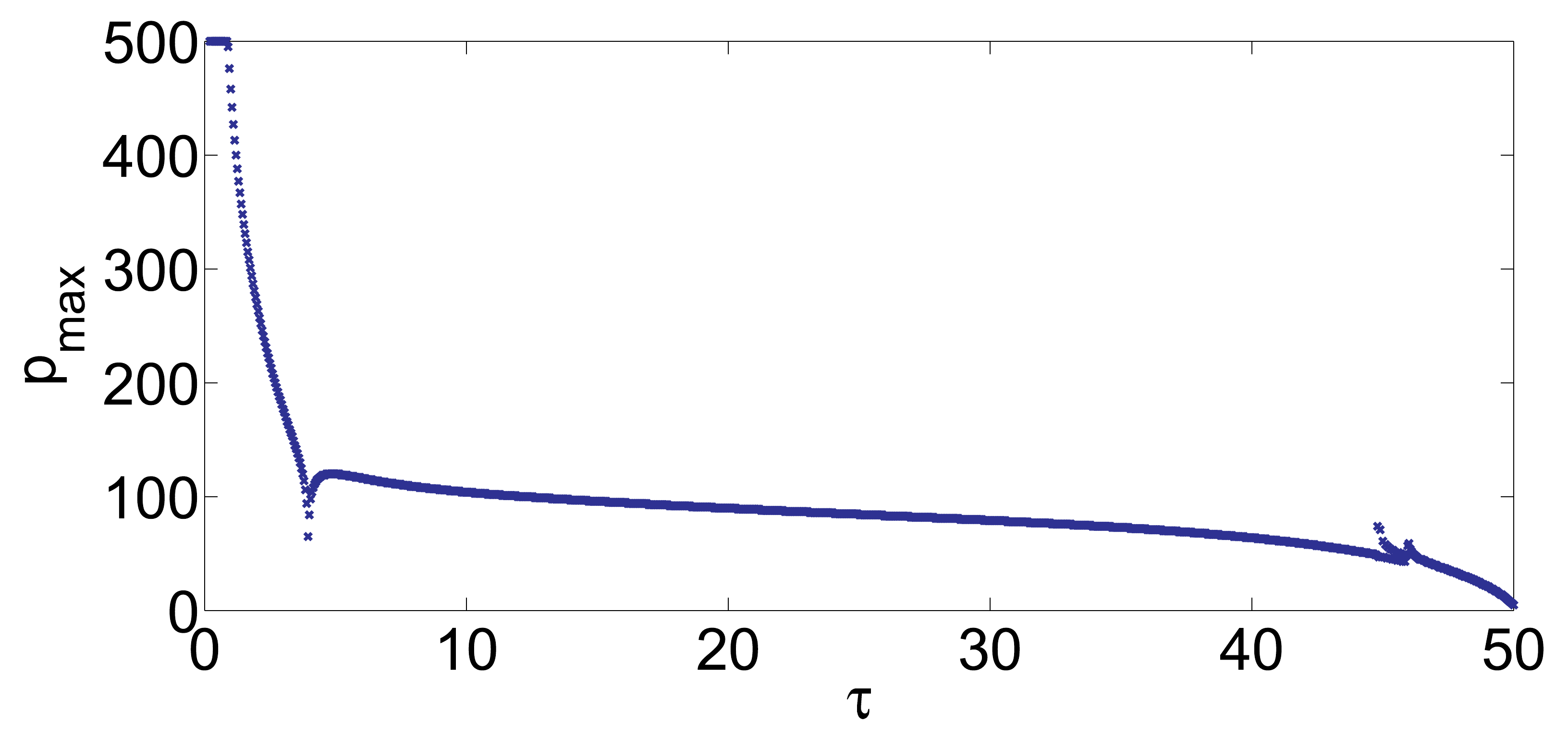}
 \caption{A number of the inner-loop iteration in the algorithm needed to achieve prescribed tolerance.} 
 \label{fig:p_errors}
\end{figure}

In Figure \ref{fig:p_errors} we show the number of maximal inner iteration steps $p_{max}$ needed for achievement of the  desired tolerance $0<toll\ll 1$.  We can observe that the algorithm requires considerably more inner iteration steps for small times $0<\tau\ll 1$ for which the time derivative of $\rho(\tau)$ is large. As for the model parameters we chose:  $T=50, \sigma = 0.2, q=0.04, r=0.06$.

\begin{figure}
 \centering
 \includegraphics[width=10truecm]{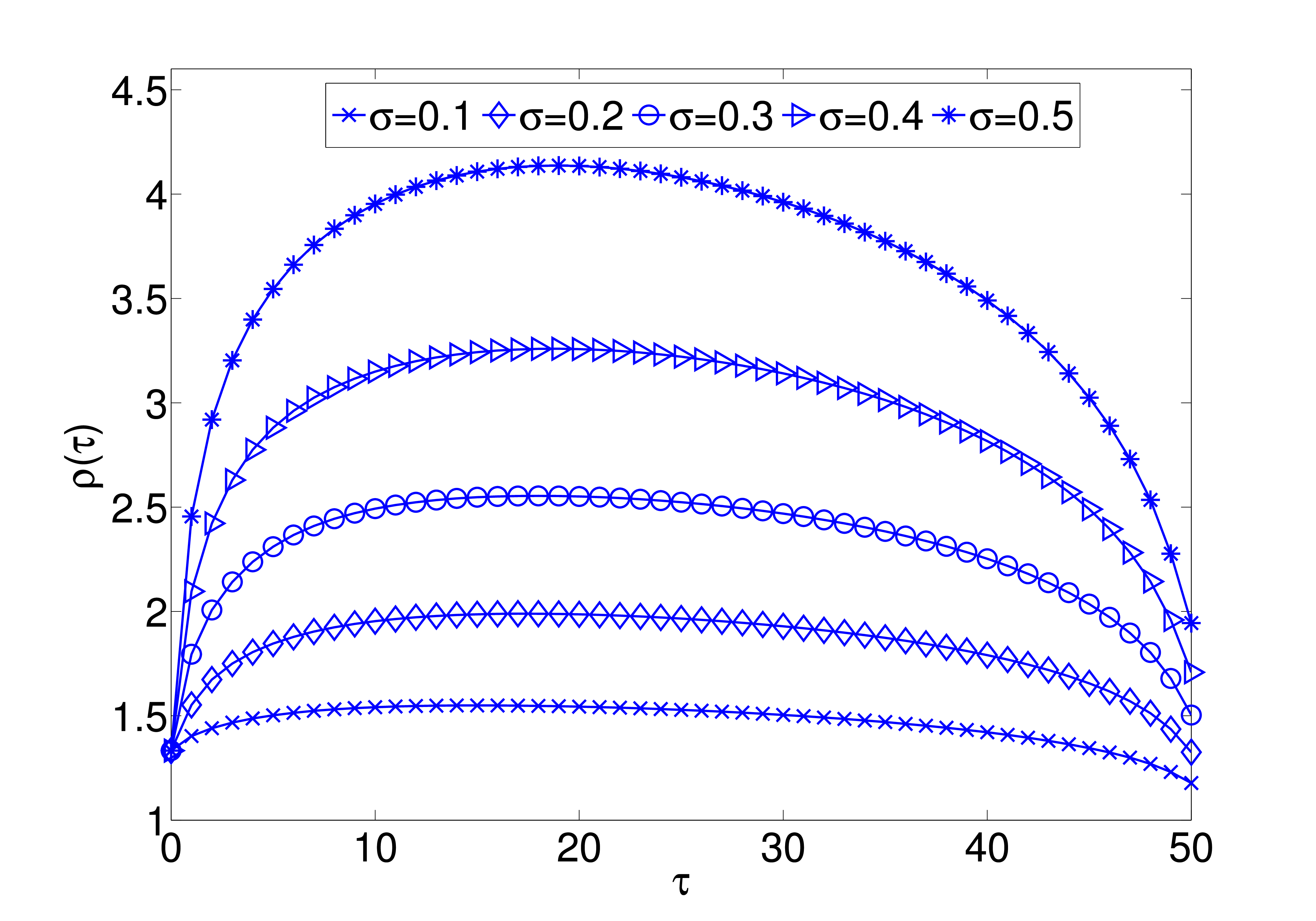}
 \includegraphics[width=10truecm]{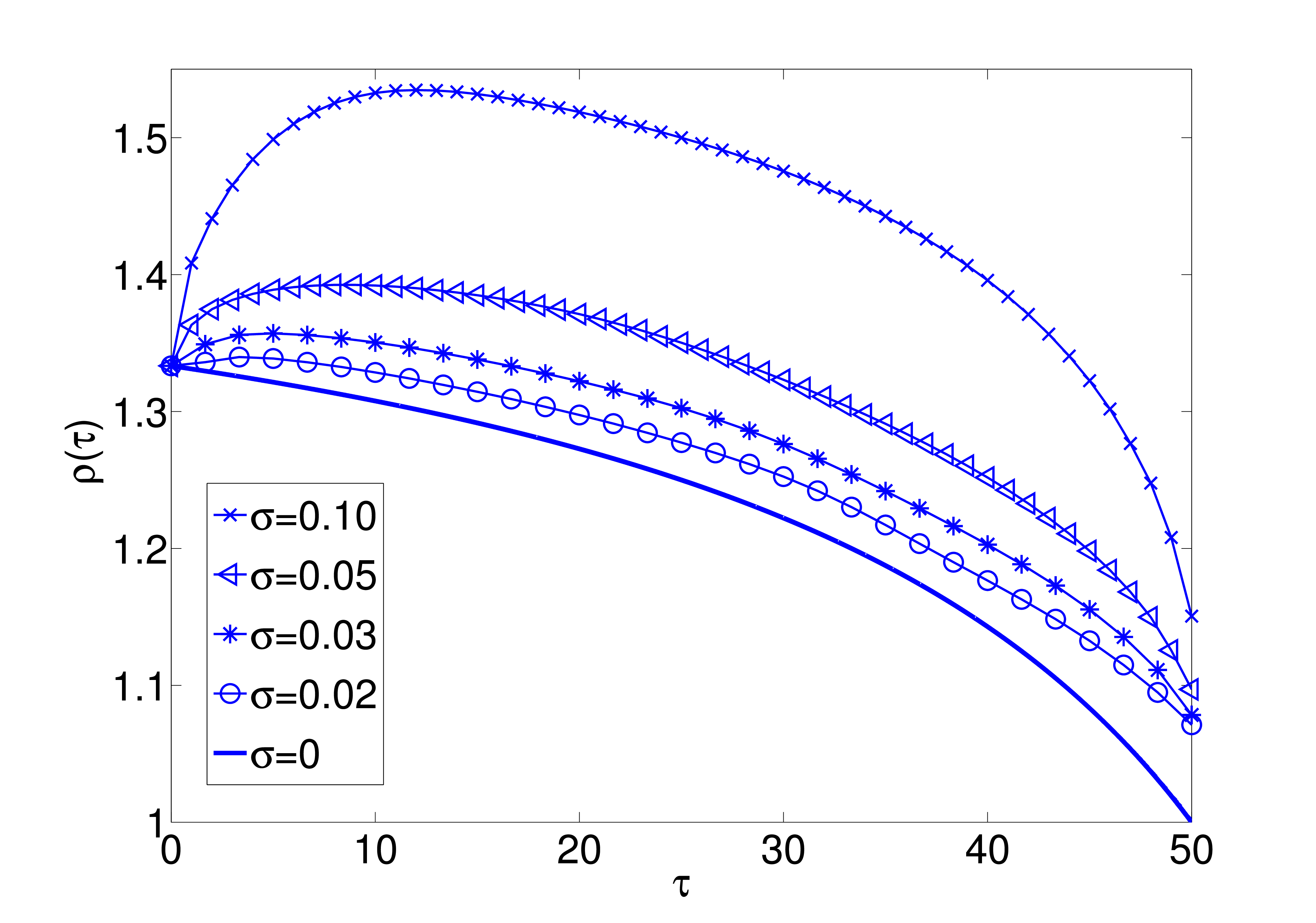}

\caption{A comparison of the free boundary position $\rho$  for large (top) and small (bottom) values of the volatility parameter $\sigma$.}
 \label{fig:sigmaCompare}
\end{figure}

In Figure \ref{fig:sigmaCompare} we plot $\rho(\tau)$ for different values of the volatility parameter $\sigma$. Other parameters are: $T=50, r=0.06, q=0.04$. It is worthwile noting that, for $\sigma \to 0$,  the underlying stochastic process for $S_t$ becomes deterministic. The option price at the time $t$ can be therefore calculated as: $S_t = S_0 e^{(r-q)t}$. Its arithmetic average is given by  $A_t = \frac 1t S_0 \frac{e^{(r-q)t} - 1}{r-q}$. In the case of $q>r$ we have $S_t < A_t$ for all $t>0$. Therefore the corresponding option price is equal to zero because it is not worth to exercise it for any $0\le t\le T$. In the case $r=q$, both $S_t=A_t$. Hence $\rho(t) \equiv 1$. Finally, in the case $r>q$ we obtain $S_t > A_t$, for all $t>0$. Since we are dealing with American style of options we exercise the option at the time $t^* =  \arg\max_{t \in [0,T]} e^{-r(T-t)}(S_t-A_t)$. Denote by $\Psi(t) = \frac{S_t-A_t}{S_0}$. Then  $t^*=\arg\max_{t \in [0,T]} S_0 e^{-rT} e^{rt} \Psi(t)$.  Now it is easy to verify that 
\[
\frac{d\Psi(t)}{dt} = \sum_{n=0}^{\infty}
   \frac{n+1}{(n+2)n!} (r-q)^{n+1} t^n > 0, \quad \hbox{for all}\ \  t > 0.
\]
Hence $\Psi(t)$ is an increasing function. As both $\Psi(t)$ and $e^{rt}$ we conclude that $t^* = T$.

In the case of an arithmetic averaged Asian call option we obtain from (\ref{eq:rovnicaPomocnaIII}) the following explicit expression for the free boundary position:
\begin{equation}
 \rho(\tau) = \max\left\{ 1, \frac{1+r (T-\tau)}{1+q(T-\tau)} \right\}, \quad \hbox{for}\ \  \sigma=0.
\label{eq:sigma0}
\end{equation}

\subsubsection{Geometric averaged floating strike call option}
\begin{figure}
 \centering
 \includegraphics[width=10truecm]{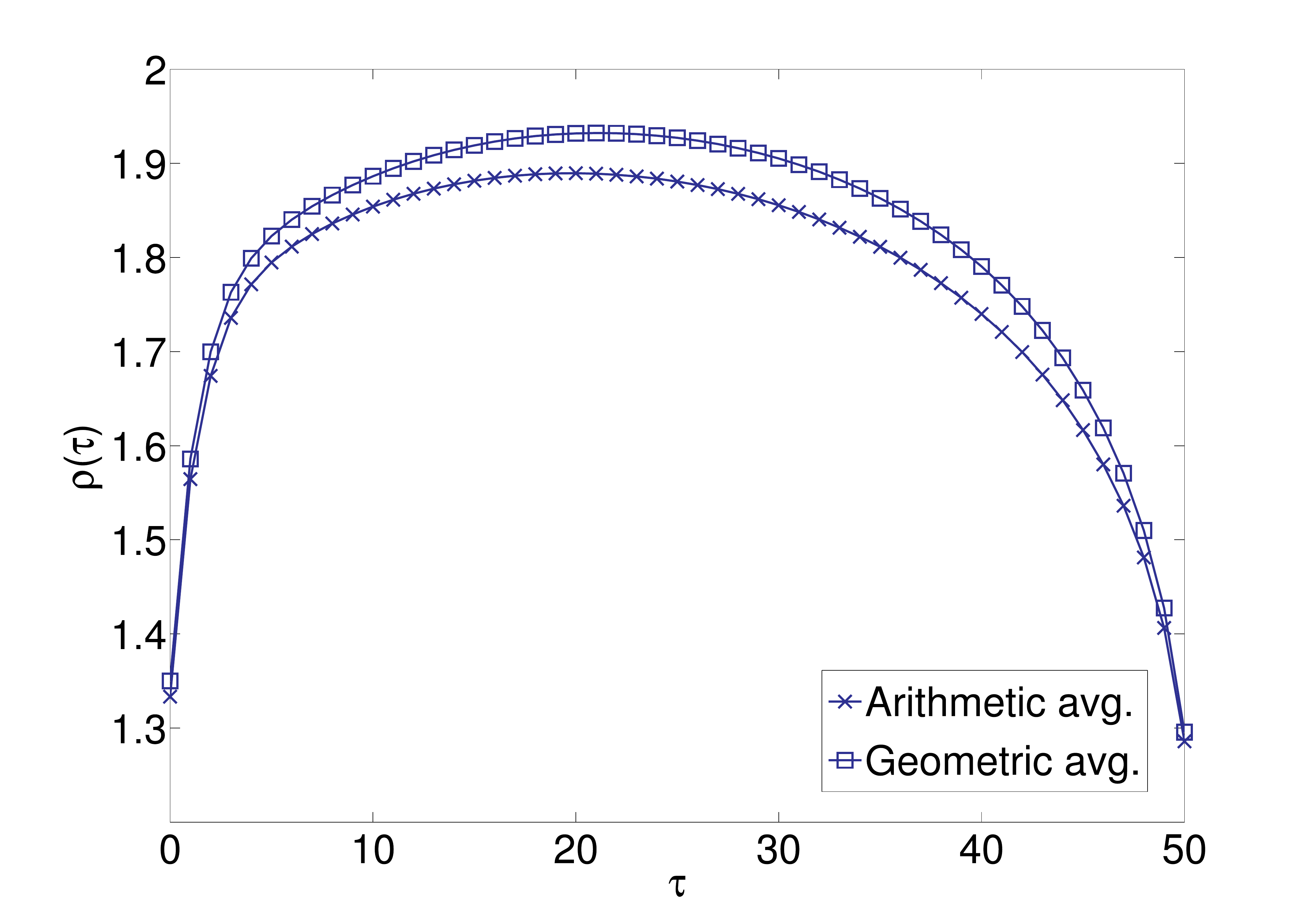}
 \caption{A comparison of the free boundary position for arithmetic and geometric averaging methods.} 
 \label{fig:aritmgeom}
\end{figure}

In Figure \ref{fig:aritmgeom} we compare the free boundary  position $\rho$ computed for the case of arithmetic and geometric averaging. Model parameters were chosen as: $T=50, \sigma = 0.2, q=0.04, r = 0.06$. Notice that $\rho^a(\tau)<\rho^g(\tau)$ for all $\tau\in [0,T]$.

\subsubsection{Weighted arithmetic averaged floating strike call option} 
\begin{figure}
\centering
\includegraphics[width=4in]{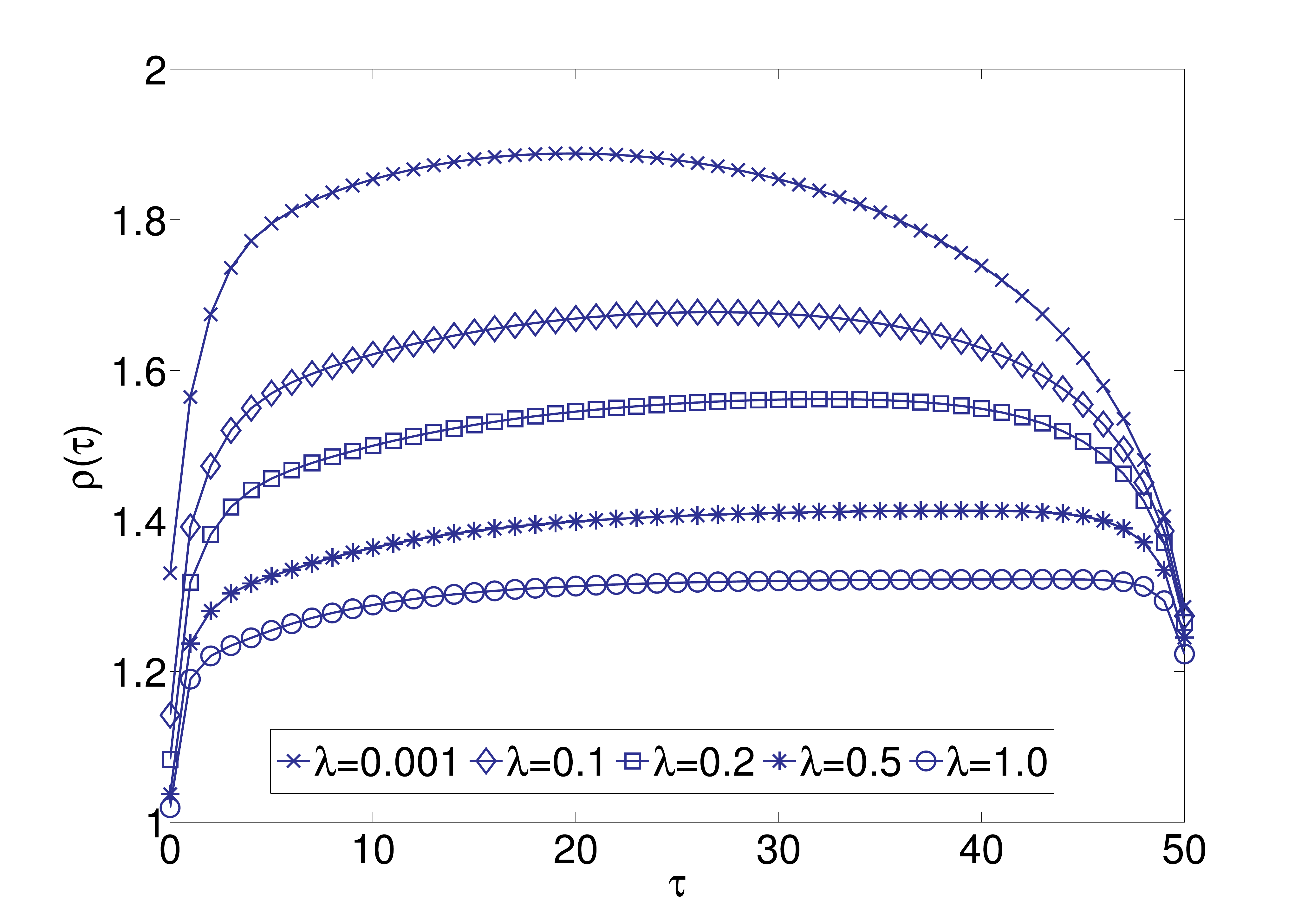}
\caption{A comparison of the free boundary position for exponentially weighted arithmetic averaged Asian options for various weight parameters $\lambda>0$. }
 \label{fig:lambdaCompare}
\end{figure}

In Figure   \ref{fig:lambdaCompare} we plot the free boundary position $\rho$ for various weight parameters $\lambda = 0.001, 0.1, 0.2, 0.5, 1$.

It is easy to verify that $\displaystyle \lim_{\lambda \to \infty} A_t^\lambda = S_t$. As a consequence we deduce that the option price is  equal to zero. This is also reason for the limit $\lim_{\lambda \to \infty} \rho_\lambda = 1$. In what follows, we will  estimate rate of convergence of $\rho_\lambda \to 1$ using the so-called experimental order of convergence. Assuming that 
\[
\Vert\rho_\lambda - \rho_\infty \Vert_\infty 
\equiv 
\max_{0\le \tau\le T} | \rho_\lambda(\tau) - \rho_\infty(\tau) | = O(\lambda^{-\alpha} ),
\] 
as $\lambda\to \infty$, we can approximate the order parameter $\alpha$ as follows:
\begin{equation}
\alpha \approx  \frac{\ln(\lambda_2) - \ln(\lambda_1) }{  \ln(\Vert\rho_{\lambda_1} - \rho_\infty \Vert_\infty)
   -\ln(\Vert\rho_{\lambda_2} - \rho_\infty \Vert_\infty)},
\end{equation}
where $\lambda_1<\lambda_2$. The results shown in Table \ref{tab:eoc_weighted} indicate  $\alpha\approx 1/3$. It means that it might be reasonable to conjecture that 
\[
\Vert\rho_\lambda - \rho_\infty \Vert_\infty = O(\lambda^{-1/3}), \quad\hbox{as}\ \lambda\to \infty.
\]

\begin{table}
 \centering
 \caption{ Experimental order of convergence $\alpha$ for the difference $\Vert\rho_\lambda - \rho_\infty \Vert_\infty = O(\lambda^{-\alpha})$ as $\lambda\to\infty$.
 }
\begin{center}
\small
\begin{tabular}{c|c|c}
  $\lambda$ & $\Vert\rho_\lambda - \rho_\infty\Vert_\infty$ 
  & $\alpha$ 
  \\  \hline \hline 
0.2 & 0.561828 & --   \\
0.5 & 0.413783 & 0.333  \\
1.0 & 0.320136 & 0.370  \\
2.0 & 0.247010 & 0.374  \\
3.0 & 0.212705 & 0.368  \\
4.0 & 0.191862 & 0.358  \\
5.0 & 0.177658 & 0.344  \\
10.0 & 0.147227 & 0.271  \\
20.0 & 0.113350 & 0.377  \\
  \hline
 \end{tabular}
\end{center}
 \label{tab:eoc_weighted}
 \end{table}

\subsection{Comparison of the early exercise boundary position for various averaging methods.}
In Remark 1 we pointed out that $\rho^{wa}(0) < \rho^a(0) < \rho^g(0)$ if $r\not= q$ and $\rho^{wa}(0) = \rho^a(0) = \rho^g(0)=1$ in the case $r=q$, where $\rho^a, \rho^{g}, \rho^{wa}$ are the free boundary position for arithmetic, geometric  and weighted arithmetic averaged Asian call options, respectively (see (\ref{nerovnosti})). This relation has been rigorously derived for $\tau=0$ only. However, it follows from results depicted in Figures \ref{fig:aritmgeom} and  \ref{fig:lambdaCompare} that it might be plausible to state the following conjecture on the comparison of early exercise boundaries for various averaging methods: 

\begin{conjecture}
Let $\rho^a, \rho^{g}, \rho^{wa}$ be the free boundary positions for arithmetic, geometric and exponentially weighted arithmetic averaged Asian call options. Then for any $0<\tau\le T$ it holds:
\[
\rho^{wa}(\tau) < \rho^a(\tau) < \rho^g(\tau).
\]
\end{conjecture}
The rigorous proof of this conjecture based on the analysis of the governing equation (\ref{finalnaPDR_prePi}) with constraint (\ref{eq:rovnicaPomocnaIII}) remains an open problem. 

\subsection{Initial early exercise boundary position}
In this section, we investigate the initial early exercise boundary position $x_f(0)$ at $t=0$. It corresponds to the value $\varrho(T)=x_f(0)$ at $\tau=T$. 

\begin{figure}
\centering
\includegraphics[width=12truecm]{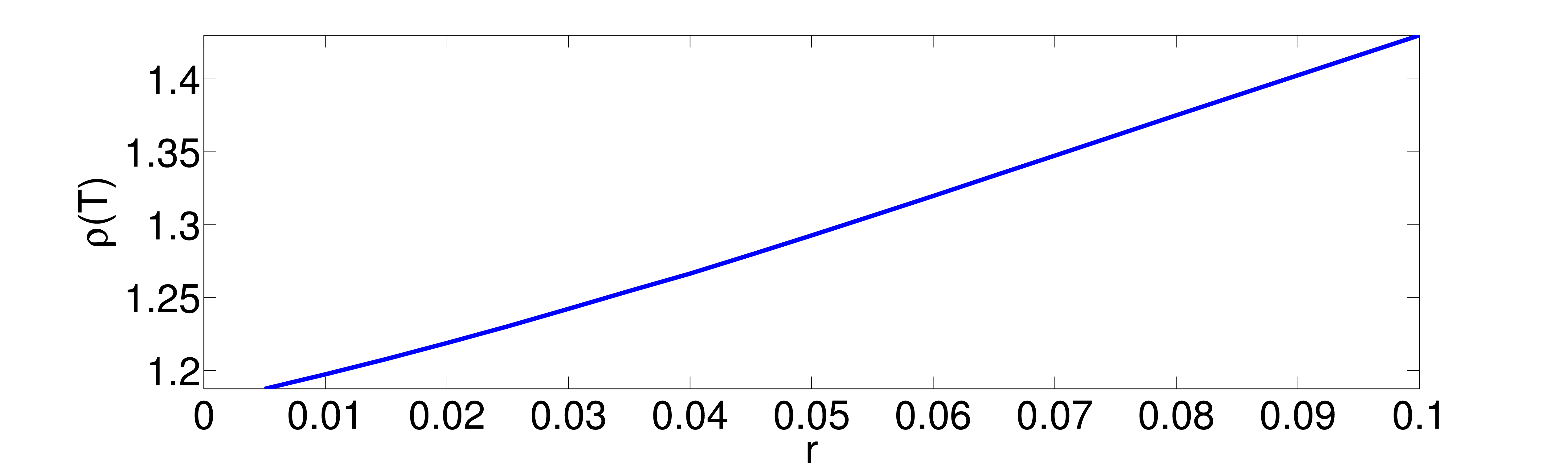}

\includegraphics[width=12truecm]{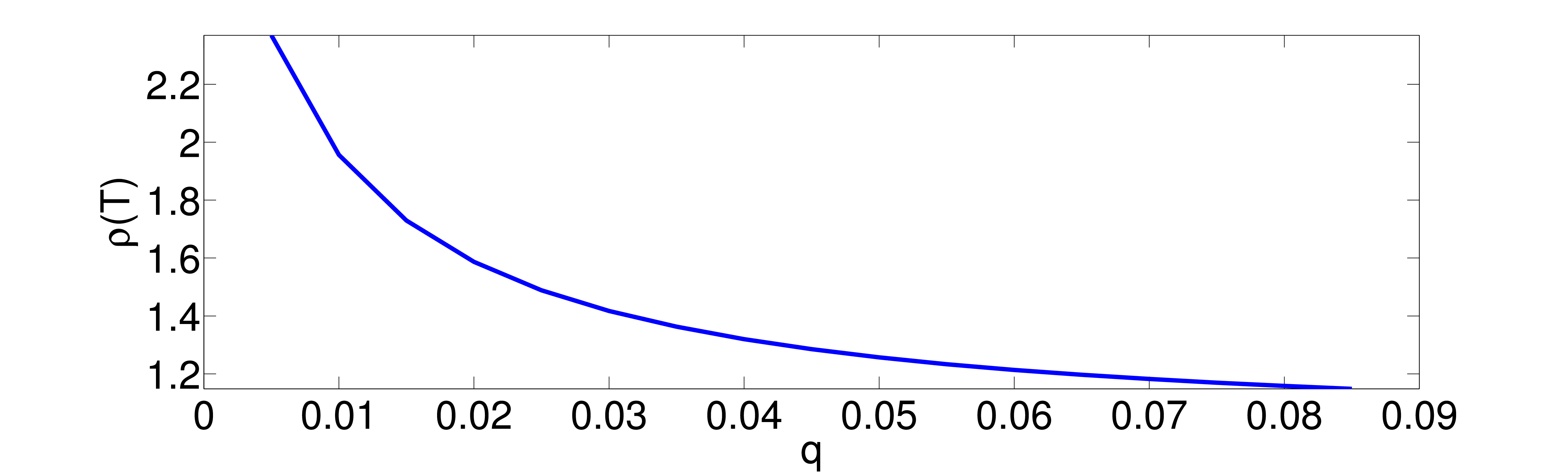}
\includegraphics[width=12truecm]{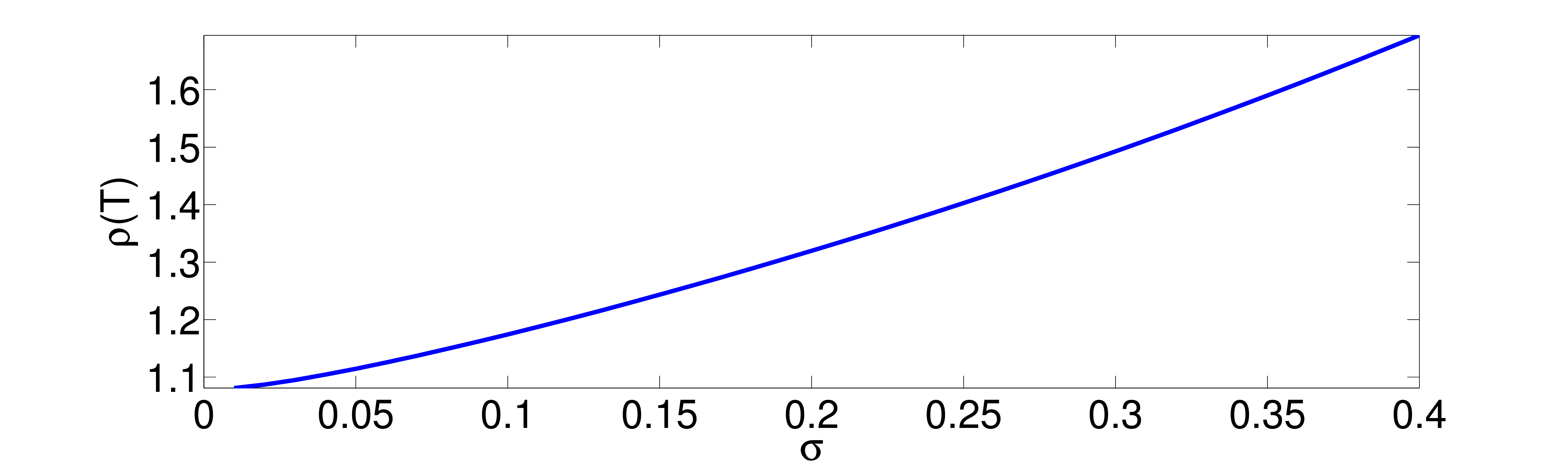}

\caption{Dependence of $\rho(T)$ on the parameter $r$ (top), $q$ (middle) and $\sigma$ (bottom).}
\label{fig:dependence_rho_T}
\end{figure}

Let us denote by $\rho(\tau) = \rho(\tau; r, q, \sigma, T)$ the free boundary position as a function of $\tau\in[0,T]$ and remaining model parameters, i.e.  $r, q, \sigma, T$. Using obvious scaling properties of the governing equation (\ref{finalnaPDR_prePi}) and the algebraic constraint  (\ref{eq:rovnicaPomocnaIII}) we can conclude that for arithmetic, geometric or weighted arithmetic average the following scaling property holds true: 
\[
\rho(\tau; r, q, \sigma, T)
= \rho\left(\frac \tau T; Tr, Tq, \sqrt T\sigma, 1\right).
\]
With regard to the previous argument, it is therefore sufficient to study dependence of $\rho(T)$ on $r,q,\sigma$ for arbitrary but fixed value of the parameter $T$. In Figure \ref{fig:dependence_rho_T} (top) we plot dependence of $\rho(T; r, q, \sigma, T)$ on $r\in(0.005, 0.1)$ for $\sigma=0.2, q=0.04$. 
In Figure \ref{fig:dependence_rho_T} (middle) we plot dependence of $\rho(T; r, q, \sigma, T)$ on $q\in(0.05, 0.1)$ for $\sigma=0.2, r=0.06$. 
Finally, in Figure \ref{fig:dependence_rho_T} (bottom) we plot dependence of $\rho(T; r, q, \sigma, T)$ on $\sigma\in(0.01, 0.4)$ for  $r=0.06, q=0.04$. In all case we chose $T=50$.

We furthermore proposed and consequently estimated the analytical formula for dependence of $\rho(T; r, q, \sigma, T)$ on $r, q, \sigma, T$. With respect to the scaling property we can fix $T=50$. Among various nonlinear estimators for the value $\rho(T; r, q, \sigma, T)$ on $r, q, \sigma, 1$ the best estimation results we achieved with the function 
\[
\varrho^{app}(T; r,q,\sigma, T) \equiv \varrho^{app}(1; T r, T q,\sqrt{T}\sigma, 1) := 1+\left(\frac{\sigma^2}{\beta_1 r+\beta_2 q}\right)^{\beta_3}+\frac{r}{q} \beta_4 
\]
with $\beta_1=-0.15064, \beta_2=7.74793, \beta_3=0.79067, \beta_4=0.09193$ with $RSS=8.6723. 10^{-4}$ where as an error indicator we chose the least square minimizer: 
\[
RSS = \sum_{j=1}^M(\varrho^{app}(T; r_j,q_j,\sigma_j, T)-\rho_j(T))^2,
\]
where $T=50$ and parameter samples $r_j,q_j,\sigma_j\in [ 0.01 , 0.11 ] \times  [ 0.01, 0.11 ] \times  [ 0.2, 0.8 ]$
were generated from $M=100$ random vectors. We denoted by $\rho_j(T)$ the numerically computed free boundary position $\rho$ at $\tau=T$ computed for the model parameters $r=r_j, q=q_j, \sigma=\sigma_j$ and time horizon $T$.

\section*{Acknowledgments}
The research was supported by ERDF-017/2009/4.1/OPVaV-CESIUK project and bilateral Slovak--Bulgarian  project APVV SK-BG-0034-08.

\end{document}